\documentclass[preprint]{aastex}
\usepackage{amsmath}
\usepackage{natbib}
\usepackage{graphicx}
\usepackage{epstopdf}
\usepackage{color}
\usepackage{rotating}
 \usepackage[T1]{fontenc}
\begin{document}
\title{Planetary chaotic zone clearing: destinations and timescales}

\author{Sarah Morrison and Renu Malhotra}
\affil{Lunar and Planetary Laboratory, The University of Arizona, Tucson, AZ 85721\\
morrison@lpl.arizona.edu, renu@lpl.arizona.edu}

\begin{abstract} 

 We investigate the orbital evolution of particles in a planet's chaotic zone 
 to determine their final destinations and their timescales of clearing.  
 There are four possible final states of chaotic particles: collision with the planet, collision with the star, escape, or bounded but non-collision orbits.  
 In our investigations, within the framework of the planar circular restricted three body problem for planet-star mass ratio $\mu$ in the range $10^{-9}$ to $10^{-1.5}$, we find no particles hitting the star.  The relative frequencies of escape and collision with the planet are not scale-free, as they depend upon the size of the planet.   
 For planet radius $R_p\ge0.001R_H$ where $R_H$ is the planet's Hill radius, we find that most chaotic zone particles collide with the planet for $\mu\lesssim10^{-5}$; particle scattering to large distances is significant only for higher mass planets.  
 For fixed ratio $R_p/R_H$, the particle clearing timescale, $T_{cl}$, has a broken power-law dependence on $\mu$.  A shallower power-law, $T_{cl}\sim \mu^{-{1/3}}$, prevails at small $\mu$ where particles are cleared primarily by collisions with the planet; a steeper power law, $T_{cl}\sim\mu^{-{3/2}}$, prevails at larger $\mu$ where scattering dominates the particle loss.   {In the limit of vanishing planet radius, we find $T_{cl}\approx0.024\mu^{-{3\over2}}$.}
 The interior and exterior boundaries of the annular zone in which chaotic particles are cleared are increasingly asymmetric about the planet's orbit for larger planet masses; the inner boundary coincides well with the classical first order resonance overlap zone, $\Delta a_{cl,int}\simeq1.2\mu^{0.28}a_p$; the outer boundary is better described by $\Delta a_{cl,ext}\simeq1.7\mu^{0.31}a_p$, where $a_p$ is the planet-star separation.
\end{abstract}

\keywords{celestial mechanics --- chaos --- planet-disk interactions --- planets and satellites: dynamical evolution and stability}

\bigskip

\section{Introduction}

Dynamical chaos arising from the overlap of orbital resonances is responsible for orbital instabilities in the solar system~\citep{Lecar:2001}.  
For the simplest case of the planar circular restricted three body problem, the overlap of first order, $p:p+1$, mean motion resonances occurs for $|p|> p_{ro}\gg1$, where $p$ is an integer, 
\begin{equation}\label{eq:pro}
p_{ro} \simeq 0.51\mu^{-{2\over7}},
\end{equation}
and $\mu\ll1$ is the planet-star mass ratio \citep{Wisdom:1980}.   This resonance overlap condition defines an approximately annular ``chaotic zone'' in the vicinity of a planet's circular orbit in which initially circular 
 test particle 
orbits are rendered strongly chaotic.  Making use of Kepler's third law with Eq.~\ref{eq:pro}, the half-width of this zone on either side of the planet's orbit is given by
\begin{equation}\label{eq:rowidth}
\Delta a_{ro} = c\,\mu^{2\over7}a_p,
\end{equation}
where $a_p$ is the semi-major axis of the planet, and $c=1.3$ is a numerical coefficient.  
An alternative analytical derivation by \citet{Malhotra:1998} found $c=1.4$.  
A numerical analysis by \citet{Duncan:1989} determined $c=1.49$ for planet mass in the range $10^{-7}<\mu<10^{-3}$.   
Although Eq.~\ref{eq:rowidth} was derived for a planet on a circular orbit, the relative insensitivity of $\Delta a_{ro}$ to planet eccentricities up to about 0.3 \citep{Quillen:2006b} makes Eq.~\ref{eq:rowidth} a powerful tool for planetary dynamics.  It is the first step in understanding the separations of long term stable planetary orbits in the solar system and in exo-solar systems.  In our solar system, the outer edge of the asteroid belt and the inner edge of the Kuiper belt are approximately coincident with this estimate of the inner and outer boundaries of the chaotic zones of Jupiter and Neptune, respectively.  {For extra-solar planetary systems, this equation has been employed to constrain the locations of planetesimal belts in some systems~\citep{Moro-Martin:2010, Schneider:2014}, and in estimating the masses of unseen planets that may be responsible for the observed gaps and edges of debris disks~\citep{Wyatt:1999, Quillen:2006, Chiang:2009, Su:2013, Rodigas:2014}. \citet{Chiang:2009} determined that Eq.~\ref{eq:rowidth}, with a coefficient of $c=2.0$, describes the cleared region of the Fomalhaut disk perturbed by an eccentric planet. \citet{Mustill:2012} and \citet{Deck:2013} have studied the dependence of the chaotic zone on the test particles' eccentricity, finding that non-zero initial eccentricities lead to wider chaotic zone widths.}

In this paper, we answer the following questions: Where do chaotic zone particles go, and how long do they take to get there? {A few previous studies have touched on these questions (e.g.  \citet{Quillen:2006, Chiang:2009, Bonsor:2011, Deck:2013, Frewen:2014}), although none have attempted a systematic study.}
To this end, we investigate the orbital evolution of particles 
in initially circular orbits in a planet's chaotic zone in the framework of the planar circular restricted three-body problem, for planet-star mass ratios in the range $10^{-9}<\mu<10^{-1.5}$ (corresponding to Pluto-mass objects to brown dwarf mass objects orbiting solar mass stars).  There are four possible final states of such particles: collision with the star, collision with the planet, escape to infinity, or bounded but non-collision orbits for infinite time.  
Using numerical integrations, we determine particle loss timescales from the chaotic zone and branching ratios for the particle final states as a function of planet mass and planet size.  
(As a practical matter, we determined the final states of particles at a finite but long time.)
 Although we refer to the massive bodies as ``star'' and ``planet'', our results are applicable to other restricted three-body astronomical contexts with small mass ratios, such as binary minor planets, star/brown-dwarf, or black-hole/star systems.  We describe our methodology in Section 2, and present our results in Section 3.  We summarize {and discuss an application of our results} in Section 4.

\section{Methodology}
The test particle orbits of interest here are strongly chaotic and must be followed accurately through close planetary encounters. 
For the numerical integration of the equations of motion, we used the Bulirsch-Stoer method~\citep{Press:1992} in the SWIFT integration package\footnote{\url{http://www.boulder.swri.edu/$\sim$hal/swift.html}}.  This method utilizes a modified midpoint method for computing integration steps and an adaptive step size for controlling numerical error.  We adopted a fractional truncation error tolerance of $10^{-10}$. { Throughout the integrations, we confirmed that there was no secular drift of the planet's semi-major axis within 1 part in $10^{16}$.}  

Within the framework of the planar circular restricted three-body problem, we adopt units such that the total mass (star and planet) is unity, the planet--star distance ($a_p$) is unity,  and the orbital period of the planet about the star is $2\pi$.  We denote by $\mu$ the ratio of the planet mass to that of the host star.
We carried out simulations for twelve values of $\log_{10}\mu$, $$\log\mu=\{-9,-8,-7,-6,-5,-4.5,-4,-3.5,-3,-2.5,-2,-1.5\}.$$
  For each of these cases, we integrated 
7840 test particles in initially circular orbits (defined as osculating circular orbits about the star) 
in an annulus about the planet's orbit.  

 For values of $\mu\leq10^{-3}$, the initial test particle orbits were uniformly spaced within each of 98 annuli in the radial range $[a_p(1-1.95\mu^{2/7}),a_p(1+1.95\mu^{2/7})]$; this is approximately 30--50\% larger than the resonance overlap zone defined by Eq.~\ref{eq:rowidth}.  In the cases of planet masses $\mu>10^{-3}$, the half-width was extended out to $3.0\mu^{2/7}a_p$ in order to more effectively determine the extent of clearing. The initial longitude of the planet was zero and the initial angular positions of the test particles were chosen uniformly over the full range, $\{0,2\pi\}$, with the exception that we excluded initial positions within $0.75R_H$ distance of the planet, where $R_H=(\mu/3)^{1/3}a_p$ is the planet's Hill radius; the excluded region is approximately the region of stable satellite orbits in the restricted three body problem.

\begin{figure}[!]
\centering
\includegraphics[scale=1]{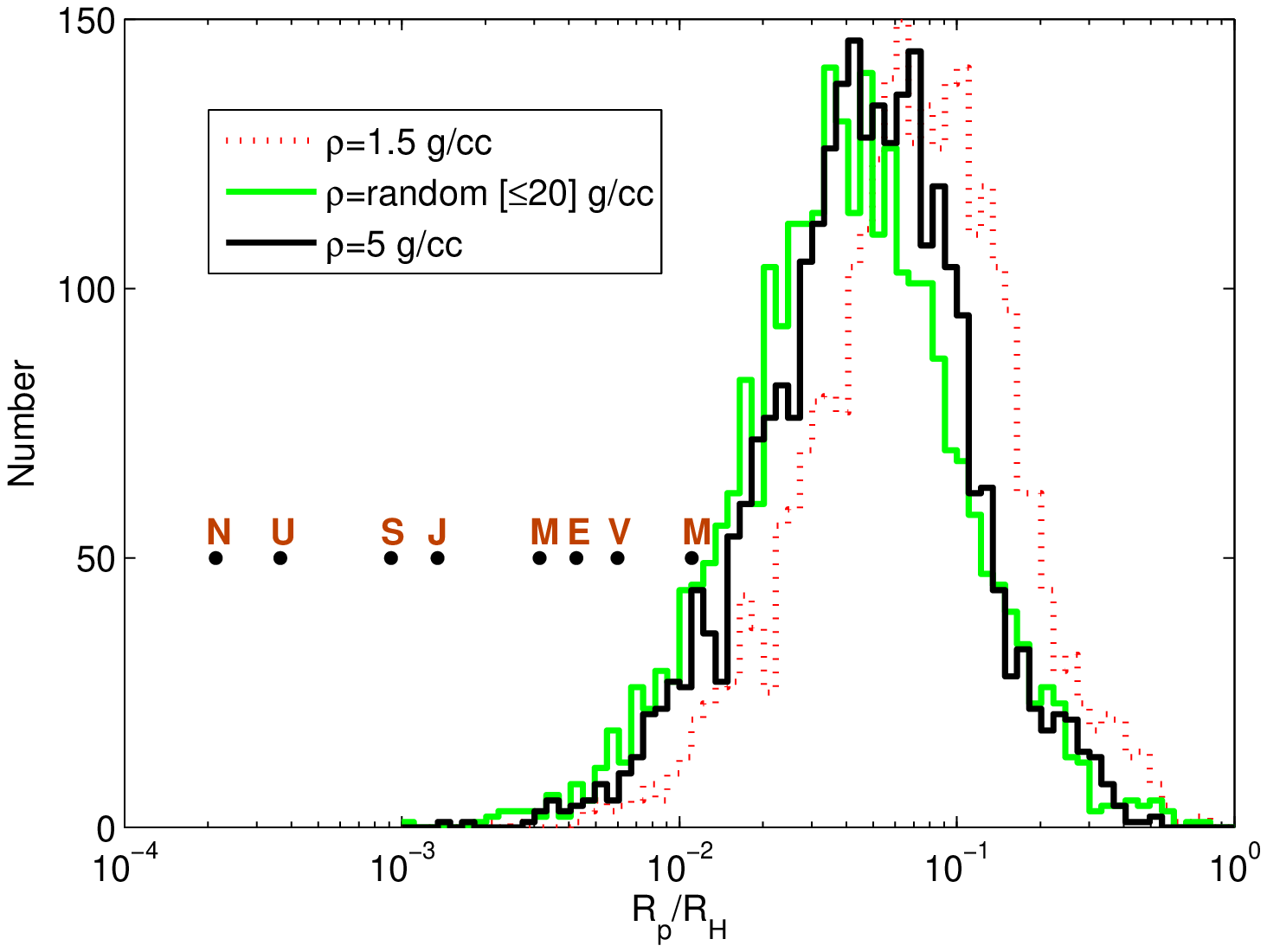}
\caption{The distribution of $R_p/R_H$, the ratio of the physical radius of a planet as a fraction of its Hill radius. The histograms represent the distribution of \emph{Kepler} planet candidates (radii from NASA Exoplanet Archive, 
 retrieved on 06 March 2014) for different assumptions of bulk density. The black dots indicate the solar system planets. }
\label{fig:rph}
\end{figure}

We integrated the test particle orbits for a time period of up to $10^8$ revolutions of the planet to determine 
 their final destinations and removal timescales. For most simulation results discussed here, integrations of $10^6$ revolutions of the planet were sufficient.  We stopped integrating a test particle under the following conditions:
\begin{enumerate}
\item {\it Particle's periastron was within $0.01a_p$}.  For $a_p=1$~AU, this distance corresponds to approximately two solar radii.  We adopted this distance as our definition for a collision with the star.  
\item {\it Particle's distance from the star exceeded $4a_p$}.  We refer to this condition as `scattered'.  Several test runs indicated that many of these particles would eventually travel to distances beyond 100$a_p$.  This choice is additionally motivated by the observation that the planets in the solar system have separations of $(2-3)a_p$; multiple planet systems amongst the known exo-planets also 
typically have smaller separations \citep{Fabrycky:2012}.
\item 
{\it Particle approached the planet within a prescribed small distance, $R_p$}.    We refer to this condition as `collision with the planet'.  We carried out three sets of simulations, one with $R_p=0.1R_H$, the second with $R_p=0.01R_H$ and the third with $R_p=0.001R_H$.

Our choice of scaling the collision radius to the planet's Hill radius is motivated by the expectation that the ratio of collisions to scatterings is governed by the square of the ratio of the surface escape velocity of the planet to the mean relative velocity of encounters, the so-called Safronov number, $\Theta=v_{esc}^2/v_{rel}^2$.  For initially circular test particle orbits in the vicinity of the planet's orbit, we expect that the typical relative velocity of encounter is on the order of the ``Hill velocity'', $v_H=(\mu/3)^{1\over3}v_p$, where $v_p$ is the planet's circular orbital velocity.  In this case, $\Theta$ is inversely proportional to the ratio of the physical radius to the Hill radius of the planet,
\begin{equation}
\Theta = {v_{esc}^2\over v_{rel}^2} \sim {2Gm_p/R_p \over (\mu/3)^{{2}\over3}Gm_*/a_p} = 6 {R_H\over R_p}.
\label{eq:Theta}
\end{equation}
The physical radius of a planet as a fraction of its Hill radius can be expressed as,
\begin{equation}
{R_p\over R_H} = \Big({3\rho_*\over\rho_p}\Big)^{1\over3}{R_*\over a_p} = 0.0042\Big({\rho_*\over\rho_\odot}{\rho_\oplus\over\rho_p}\Big)^{1\over3}{R_*\over R_\odot}\Big({a_p\over 1~\hbox{AU}}\Big)^{-1},
\label{eq:Rp}
\end{equation}
where $\rho_p,\rho_*$ are the mean density of the planet and star, respectively,  $\rho_\oplus$ is the mean density of Earth, $R_*$ is the stellar radius, and $\rho_\odot$ and $R_\odot$ are the mean solar density and radius.  This fraction is only weakly dependent on planet density, and is mainly dependent on planet-star separation.
The currently known planets in the Galaxy have $R_p/R_H$ concentrated in the range 0.01--0.1, as shown in Figure~\ref{fig:rph}, since the majority of planets are discovered at small orbital radii by means of radial velocity and transit observations.  The solar system planets have smaller values (also indicated in Figure~\ref{fig:rph}), as do directly imaged planets, due to their greater distances from their host star.

\end{enumerate}
Particles that did not meet the above conditions over the duration of the integrations are referred to as `survivors'.

 We note that there is one known conserved quantity, the Jacobi integral~\citep{hill:1878}, that constrains the evolution of test particles in the circular restricted three body problem.  For the planar case, the Jacobi integral is given by
\begin{equation}\label{eq:Jacobi}
J =-v^2 + 2(x\dot y-\dot x y)+2\left(\frac{\mu_1}{r_1}+\frac{\mu_2}{r_2}\right),
\end{equation}
where $\mu_1=(1+\mu)^{-1}, \mu_2=\mu(1+\mu)^{-1}$, and $r_1$ and $r_2$ are the distances of the test particle from the star and the planet, respectively, and $(x,y)$ and $(\dot x,\dot y)$ are the particle's position and velocity vector in an inertial reference frame with origin at the barycenter of the star and the planet.   
A particle's range in the $x$--$y$ plane is bounded to regions where $\dot x^2+\dot y^2 \ge 0$, therefore Eq.~\ref{eq:Jacobi} (with $\dot x=\dot y = 0$, defines the boundary of this region, the so-called zero-velocity curve.  It is pertinent to note that the zero-velocity curves for most of our initial conditions of particles within the chaotic zone allow all possible end-states, including collision with the star, collision with the planet, scatter to infinity, or possibly long-lived bounded orbits that may be either chaotic or quasi-periodic; i.e., the value of $J$ alone is generally not indicative of the final destinations of our chaotic zone particles.  A small minority of initial conditions, near the inner and outer edges of the chaotic zone annulus, have values of $J$ that are bounded to remain entirely interior or entirely exterior to the planet's orbit.

In addition to meeting the truncation error tolerance in the numerical integrations, we monitored the value of the Jacobi integral, $J$, for each test particle.

For the range of initial conditions of interest, $J$ is close to 3, and $|J-3|$ is small, on the order of $\mu^{4/7}$. 
We found that most of the chaotic zone test particles preserved their values of $J$ to better than a part in $10^5$, but high numerical accuracy of the preservation of the Jacobi integral was sometimes more difficult to achieve than meeting the local truncation error tolerance.  Unsurprisingly, the cases of small values of $R_p/R_H$ pose a particularly stiff computational challenge.   We identified numerical solutions as inaccurate if, over the course of the integrations, the fractional variation of $J$ exceeded $2\mu^{4/7}/(3+2\mu^{4/7})$. Simulations with higher planet masses and lower values of $R_p/R_H$ had larger numbers of these cases. For all cases of $R_p/R_H=0.1, 0.01$ and $\mu\leq10^{-2}$ that we simulated, we found that no more than $4\%$ of our test particles failed this Jacobi conservation criterion; the case of $\mu=10^{-4.5}$ for $R_p=0.001R_H$ had the highest fraction, 27\%, of the test particles failing our Jacobi conservation criterion. 
 We found no obvious pattern in the initial conditions of these particles. 
These cases were excluded from the analysis that we report in the following sections. 
Discarding these cases has little effect on the reported results.

\section{Results}

\subsection{Final destinations of chaotic zone particles}\label{s:destinations}

\begin{figure}[!]
\centerline{\includegraphics{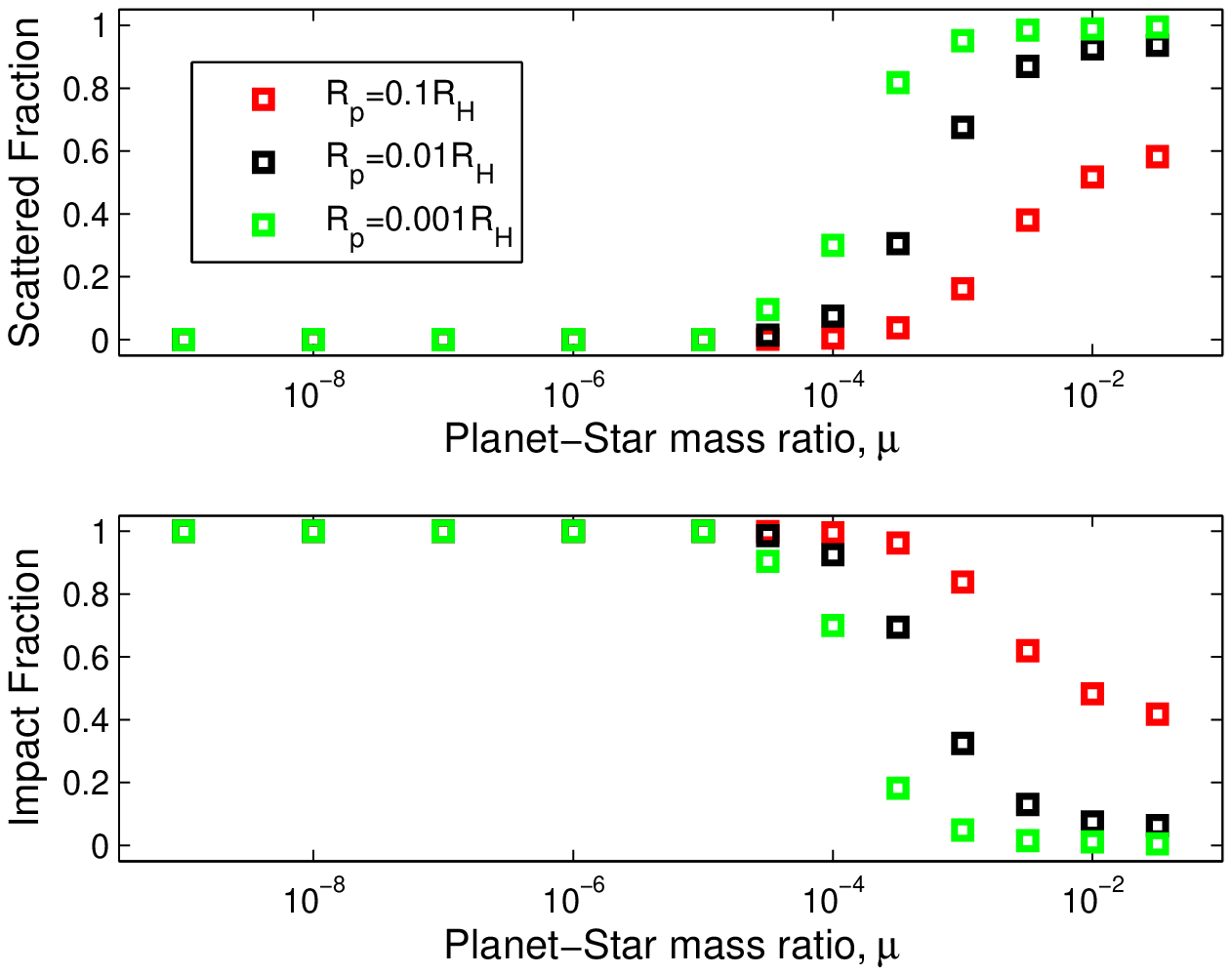}}
\caption{Fraction of non-surviving particles in the chaotic zone that are scattered (top) or impact the planet (bottom), as a function of $\mu$, for fixed $R_p/R_H$.} 
\label{fig:fr}
\end{figure}

\begin{figure}[!]
\centerline{\includegraphics{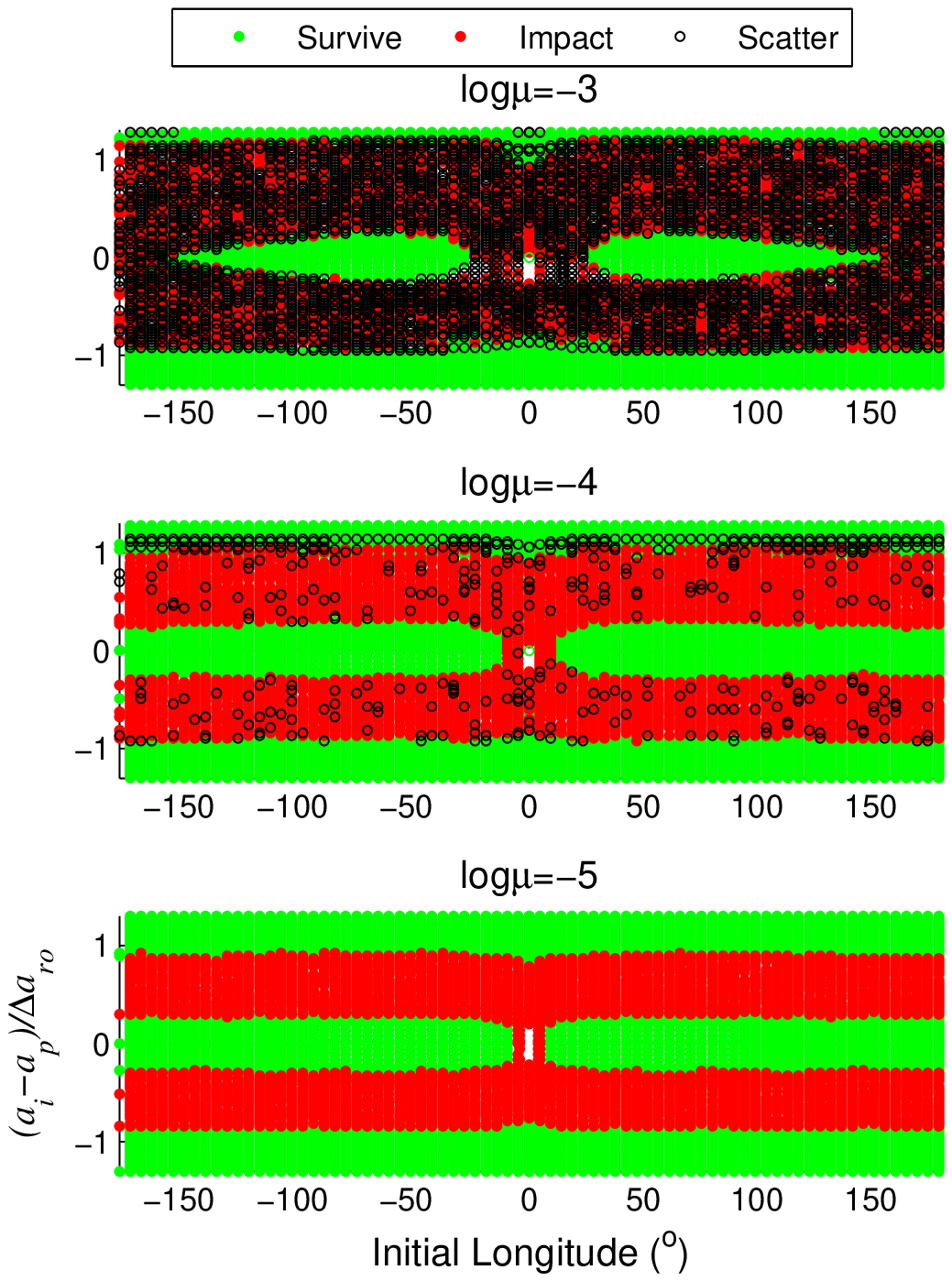}}
\caption{ Final destination of chaotic zone particles as a function of particle initial conditions.   Black points indicate scattering, red indicates impact with the planet, and green indicate survivors to the end of the simulation. We show results for three values of $\mu$, for the case of $R_p=0.01R_H$. For clarity, we scale the initial semimajor axis separation from the planet, $a_i-a_p$, to the half-width of the chaotic zone, defined as $\Delta a_{ro}=1.5\mu^{2/7}a_p$. { White patches indicate particles that were excluded due to poor conservation of the Jacobi constant.} }
\label{fig:br}
\end{figure}

Overall, we find that particles do not hit the star in any of our simulations. 
Scattering to large distances is the most frequent outcome at larger planet masses;  impact with the planet becomes  relatively more frequent at larger planet sizes.  We find that impact with the planet was the sole fate of non-surviving particles for planetary masses $\mu\lesssim10^{-5}$ and $R_p/R_H\ge0.001$. 
Scattering is the dominant loss mechanism when $\mu\gtrsim10^{-3.5}$, $10^{-3}$, and $10^{-2}$, for $R_p/R_H=$0.001, 0.01, and 0.1, respectively.  
Figure~\ref{fig:fr} plots, as a function of $\mu$, the fraction of non-surviving particles that were scattered or that collided with the planet, for each case of $R_p/R_H$.

The final destination of test particles as a function of initial semimajor axis and longitude are shown in Figure~\ref{fig:br}, for three representative values of $\mu$ for the case of $R_p=0.01R_H$.  (Figures for all other simulated cases of $\mu$ and $R_p/R_H$ are available by request from the authors.)
We observe a noticeable asymmetry in Figure~\ref{fig:br} (see also Figure~\ref{fig:tvaall}) in the fates of the interior versus exterior particles with initial distance from the planet,  especially for larger planet masses.  The range of initial semimajor axis of particles that scatter or collide with the planet is smaller and closer to the planet's orbit in the case of interior particles, and is relatively larger and extends farther from the  planet's orbit in the case of exterior particles.

In Figure~\ref{fig:br}, we also observe that most particles with initial orbits such that $|a_i-a_p| \lesssim R_H$ are survivors; a minor fraction of these collide with the planet.  This surviving population is associated with the triangular Lagrange points, L$_4$ and L$_5$.  These particles are stable over the duration of our integrations, and survive in low eccentricity orbits ($e\ll0.05$) either as Trojan-type orbits librating about L$_4$ or L$_5$ (analogous to the Trojan asteroids of Jupiter), or as ``horseshoe'' orbits librating about both L$_4$ and L$_5$.  For $10^{-9}\le\mu\leq10^{-4}$, we find that  these ``co-orbital'' survivors 
 span a radial range of $20-30\%$ within the nominal chaotic zone annulus, [$a_p(1-1.5\mu^{2\over7}),a_p(1+1.5\mu^{2\over7})$]. {Larger ranges of initial longitudes in the co-orbital region are stable for lower planet masses.}  The stable co-orbital region decreases significantly for planet masses $\mu>10^{-4}$, and it vanishes for $\mu\ge10^{-2}$.  This behavior is likely related to the well known linear instability of the triangular Lagrange points, L$_4$ and L$_5$, for $\mu\gtrsim10^{-1.4}$~\citep{murray:1999SSD}.

Survivors are also found near the outer edges of the chaotic zone.  A few of these particles survive in nearly circular orbits, in libration in a first order mean motion resonance, and exhibit small ($<1\%$) fractional changes in their semimajor axes.  
But, more typically, the survivors near the boundaries of the chaotic zone are found on eccentric orbits, { with eccentricities up to $\sim0.6$. }  
 This population is larger for the chaotic zones of lower mass planets.  For our smallest planet mass,  $\mu=10^{-9}$, the outer $\sim45\%$ of the radial extent of the chaotic zone survives on eccentric orbits for more than $10^6$ revolutions of the planet.  { However, these particles spend insignificant time in the ``cleared zone'' defined below.}

\subsection{Cleared zone}\label{s:clearedzone}

\begin{figure}[!]
\centering
\includegraphics[scale=1]{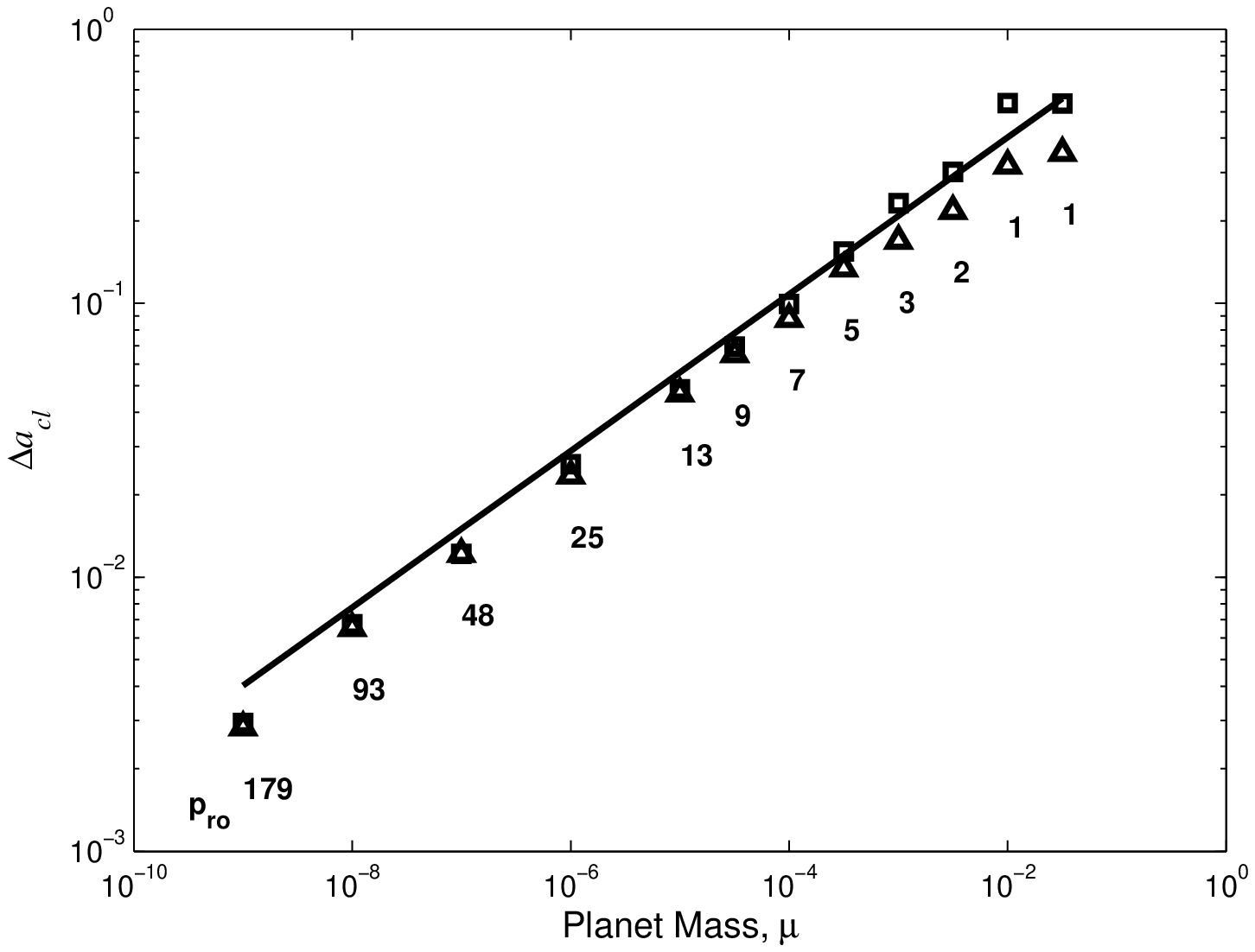}
\caption{Interior (triangles) and exterior (squares) cleared regions,  $\Delta a_{cl}$, for initially circular orbits near a planet with $R_p=0.01R_H$.  Numbers under each point correspond to {the values of $p_{ro}$ in Eq.~\ref{eq:pro}, the analytical estimate of the most distant overlapping $p:p+1$ resonance. The straight line indicates the chaotic zone width from the first order resonance overlap criterion (Eq.~\ref{eq:rowidth}), with c=1.5.}}
\label{fig:reso}
\end{figure}

 In previous studies, the boundaries of the chaotic zone have generally been defined in terms of the initial conditions.  For practical astronomical applications (for example, gaps in protoplanetary or debris disks), it is of interest to define the extent of the ``cleared zone'', the zone in which the density of surviving particles is nearly vanishing.   
 In order to compute the radial boundaries of this zone, we adopted the following procedure.
We first computed the time-averaged astro-centric distance of the surviving particles in their final osculating orbits, i.e., $\langle r\rangle = a_f(1+{1\over2}e_f^2)$.  We binned these values into 98 bins in the range $a_p(1-1.95\mu^{2\over7})$ to $a_p(1+1.95\mu^{2\over7})$; these are the same bins that we adopted for the osculating semimajor axes of the test particle initial conditions.  We computed the survivor fraction (the ratio of the number of survivors to the initial number) in each annular bin.  We then determined the interior and exterior boundaries of the cleared zone by finding the most distant annulus on either side of the planet that has a survivor fraction less than 5\%.  
Figure~\ref{fig:reso} shows these inner and outer boundaries of the cleared zone as a function of $\mu$, for the case of $R_p/R_H=0.01$. 

We observe a noticeable asymmetry in the interior and exterior cleared zones, which increases with increasing $\mu$.  
For each case of $R_p/R_H$, we computed a least-squares fit to a power law of the form 
\begin{equation}\label{eq:clwidth}
\Delta a_{cl} = C\,\mu^{\beta}a_p,
\end{equation}
to the interior and exterior boundaries of the cleared zone.
The values of the parameters $C$ and $\beta$ are given in Table~\ref{tbl:clzone}.  
We find, unsurprisingly, that both the power law index and the numerical coefficient have little dependence on the value of $R_p/R_H$.  The cleared zone boundary interior to the planet's orbit mimics the first order resonance overlap criterion, Eq.~\ref{eq:rowidth}, with $c\simeq 1.2$.   The exterior boundary has a slightly steeper power law index as well as a larger numerical coefficient; it is better described by $\Delta a_{cl,ext} = 1.7 \mu^{0.31}a_p$.  {This asymmetry, which increases with increasing $\mu$, arises because the locations and strengths of interior and exterior overlapping first order mean motion resonances are not symmetric.}

\begin{table}[h]
\caption{Best-fit power law function for the interior and exterior boundaries of the cleared zones.}
\label{tbl:clzone}
\begin{center}\begin{tabular}{l l l l l r r}
\cline{2-7}
                                & \multicolumn{3}{l}{Interior Boundary} & \multicolumn{3}{r}{Exterior Boundary} \\ \hline\hline
\multicolumn{1}{l}{$R_p/R_H$} & $\beta$          & $C$   &     &  & $\beta$          & $C$              \\ \hline
\multicolumn{1}{l}{0.001}     & 0.28$\pm$0.01    & 1.17$\pm$0.13 &  &     &   0.31$\pm$0.01    & 1.76$\pm$0.28    \\
\multicolumn{1}{l}{0.01}      & 0.28$\pm$0.01    & 1.19$\pm$0.10  &    &  &   0.30$\pm$0.01    & 1.68$\pm$0.24    \\ 
\multicolumn{1}{l}{0.1}       & 0.29$\pm$0.01    & 1.31$\pm$0.22  &  &  &   0.31$\pm$0.01    & 1.79$\pm$0.18    \\ \hline
\end{tabular}\end{center}
\end{table}

It is also interesting to examine the critical integer value, $p_{ro}$ (Eq.~\ref{eq:pro}), for which all $p:p+1$ 
 resonances having $|p|\geq p_{ro}$ are overlapping for a given value of $\mu$.
The nearest integer defined by Eq.~\ref{eq:pro} is indicated in Figure~\ref{fig:reso}, adjacent to the values of $\mu$ investigated in our numerical simulations. We observe that for the larger values of $\mu$, $p_{ro}$ moves into the range of single digits.  In this range, where the first order resonances are sparse, we can expect less smoothness in the relationship between $\Delta a_{ro}$ and $\mu$, as the overlap of neighboring resonances occurs at discrete intervals of $\mu$.  Indeed, for $\mu\gtrsim10^{-2}$, we reach critical integer value $p_{ro}=1$; this means that all possible first order mean motion resonances are overlapping.  (We note that, formally, the assumption of $|p|\gg1$ in the derivation of Eq.~\ref{eq:pro} fails for $\mu\gtrsim10^{-3}$.)  This could explain the diminished mass dependence on the 
 size of the cleared zone for these mass ratios seen in Figure \ref{fig:reso}.

\subsection{Clearing Timescales}\label{s:timescales}

\begin{figure}
\centerline{\includegraphics[scale=1]{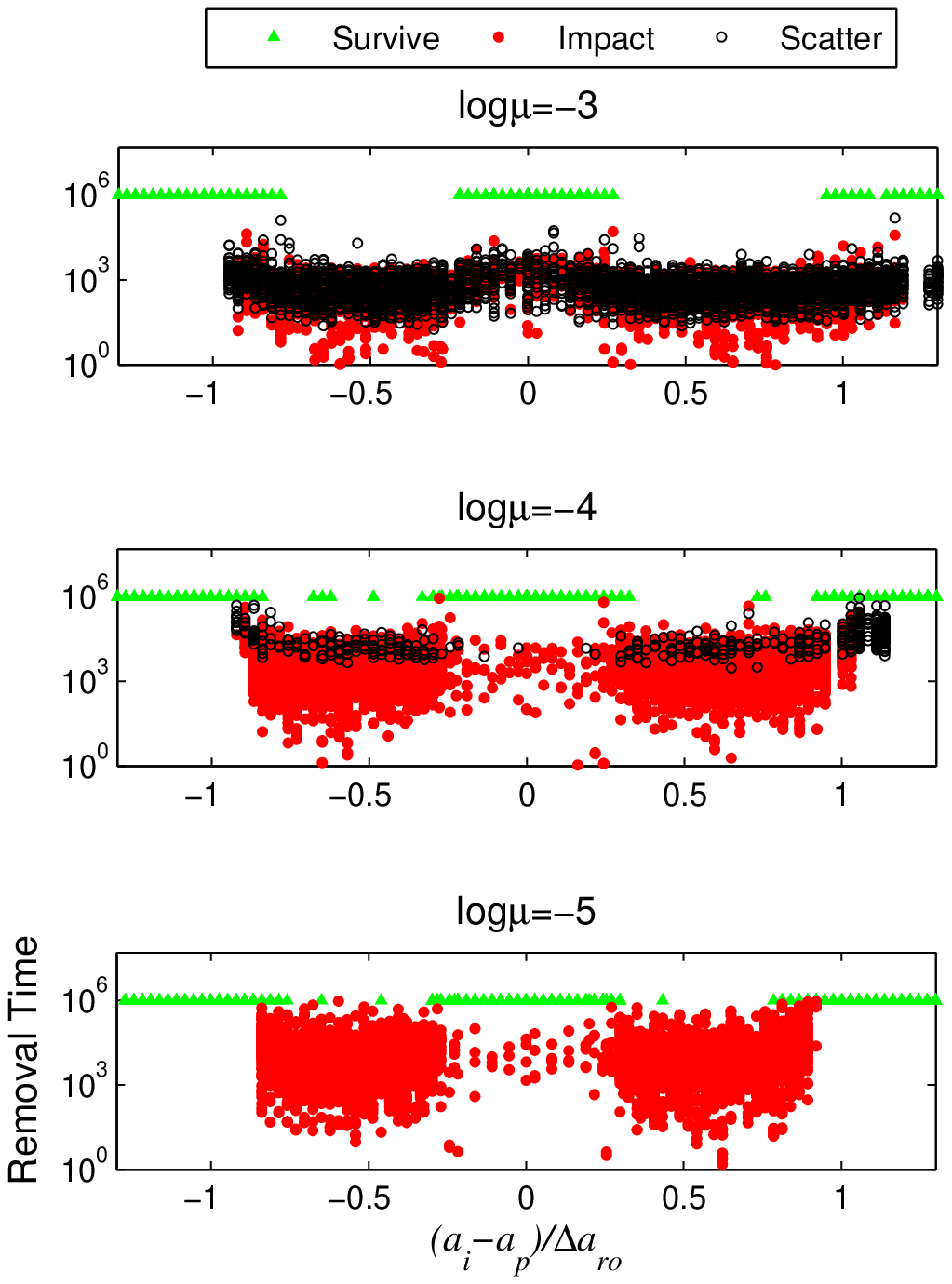}}
\caption{Removal time as a function of initial semimajor axis for three values of $\mu$, for the case of $R_p=0.01R_H$. Black points indicate test particles that were removed via scattering and red points indicate test particles removed via impact with the planet. Green points indicate survivors at the end of the simulation.} 
\label{fig:tvaall}
\end{figure}

The timescale for particles to be cleared from the chaotic zone generally decreases with increasing $\mu$.   There is also a noticeable difference in 
the time required to remove test particles via collision with the planet or via scattering.  Figure~\ref{fig:tvaall} plots the removal time versus initial semimajor axis; the color of the points indicates the removal mechanism (scattering or collision with planet); the survivors are assigned a minimum lifetime equal to the duration of the simulation.  
As mentioned above, we find that particle removal via scattering to large distances occurs only for planet masses $\mu\geq10^{-4.5}$, and it occurs throughout the chaotic zone. 
The scattered particles have longer removal times than those removed via collision with the planet. 
It is evident in Figure~\ref{fig:tvaall} that the range of initial conditions for which particles are cleared is asymmetric between the interior and exterior zones; however, their clearing timescales are similar.

\begin{sidewaysfigure}

\centerline{\includegraphics[scale=0.5]{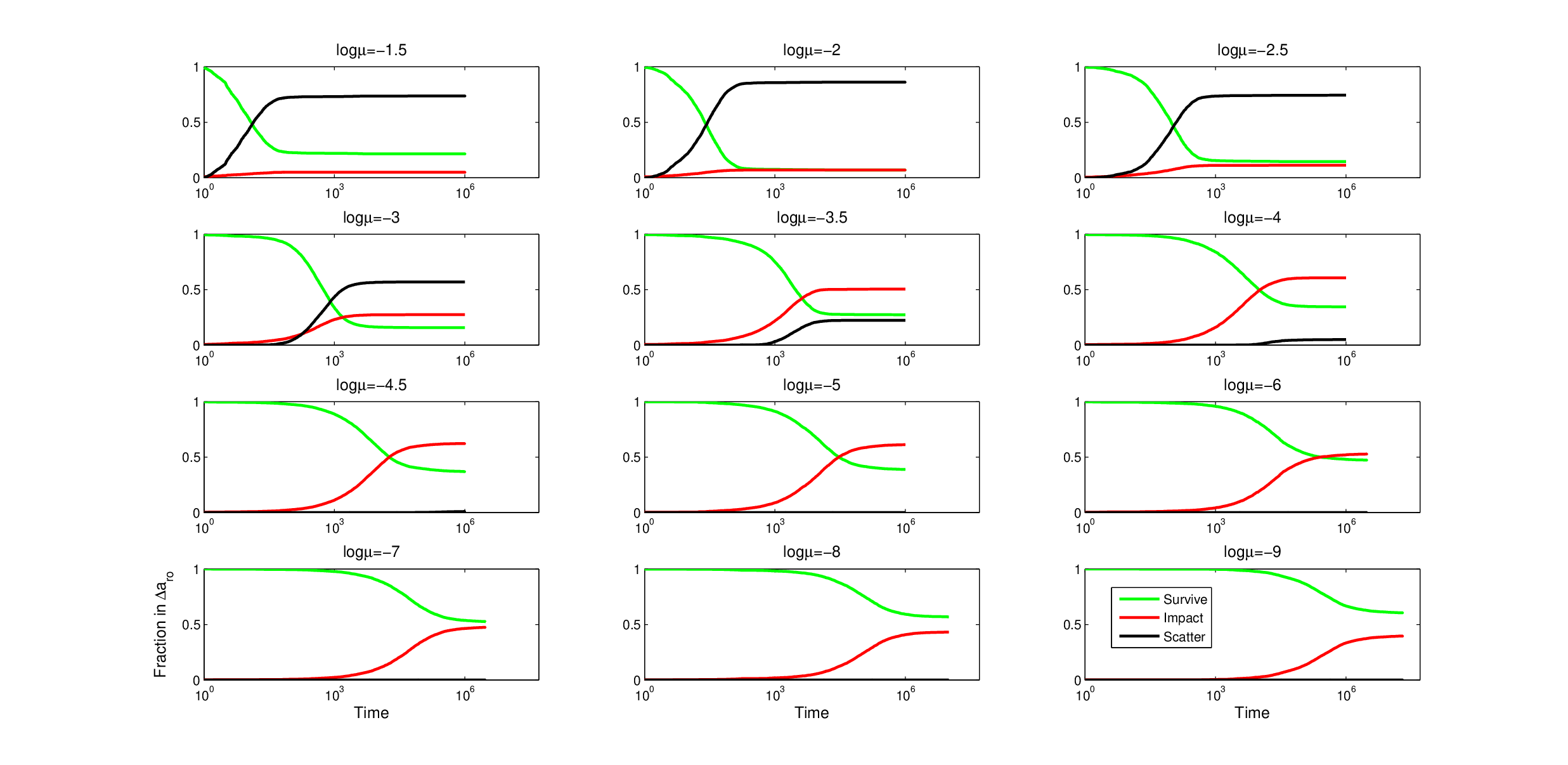}}
\caption{
Time evolution of clearing within the chaotic zone, for planet masses ranging from $\mu=10^{-9}$ to $10^{-1.5}$, for the case of $R_p=0.01R_H$. Green lines indicate the survival fraction, red lines indicate the fraction of test particles that impact the planet, and black lines indicate the scattered fraction.} 
\label{fig:tevoall}
\end{sidewaysfigure}

\begin{figure}
\centering
\includegraphics[scale=1.0]{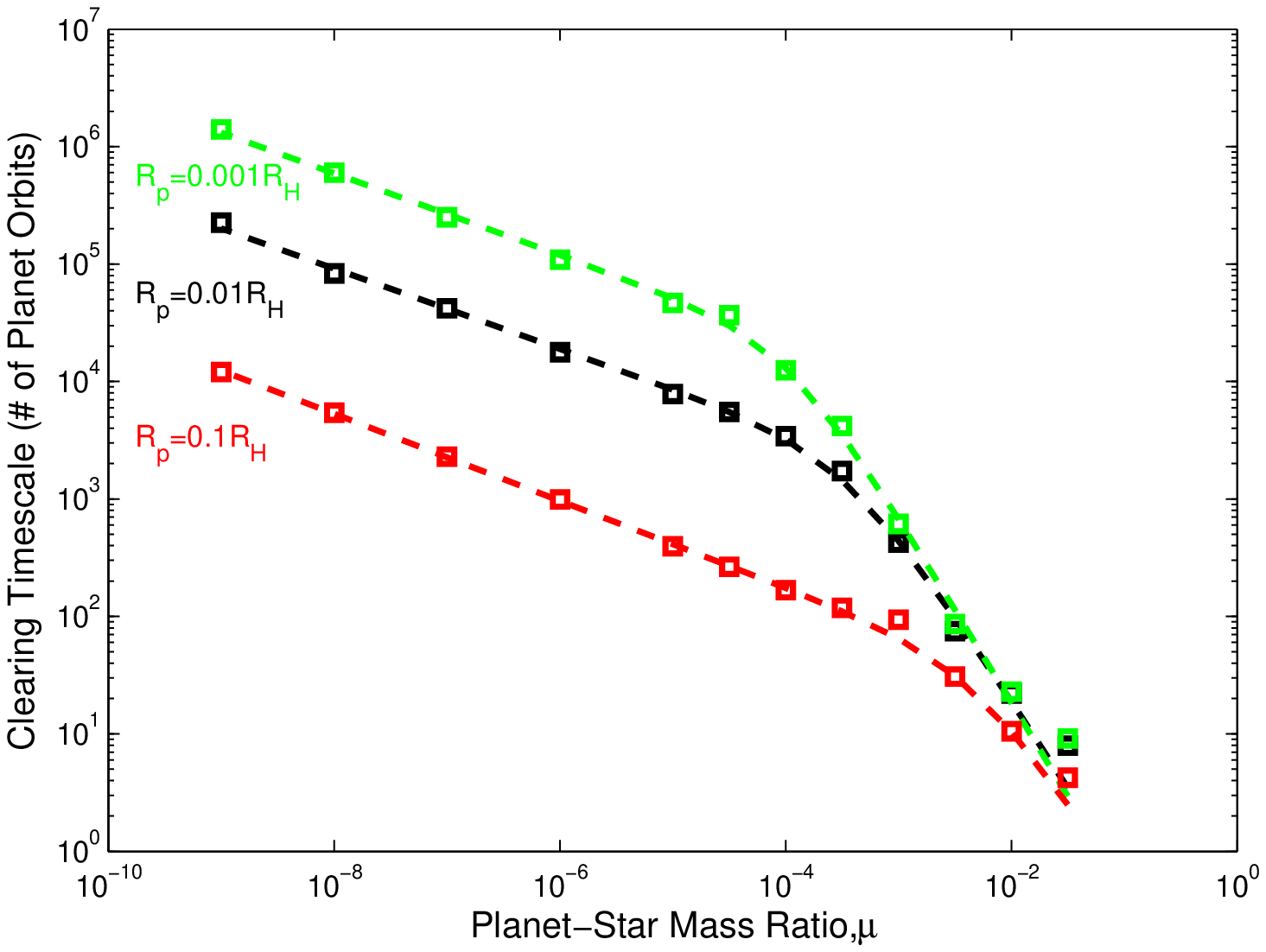}
\caption{ Clearing timescale (in units of the planet's orbit period) as a function of planet mass, for the three cases of $R_p/R_H$.   The best-fit double power law models (Eq.~\ref{eq:Tcl}, Table~\ref{tbl:Tcl}) are shown as the dashed lines. } 
\label{fig:Tcl}
\end{figure}

The time evolution of the test particle population in the combined interior and exterior 
 cleared zone is shown in Figure~\ref{fig:tevoall} 
 for all the values of $\mu$ that we investigated, for the case of $R_p=0.01R_H$. 
We plot the survivor fraction, the fraction that collide with the planet and the fraction that scatter, all as a function of time.  
We find that the `survivor' fraction decays and eventually stabilizes to a final (non-zero) value over a characteristic time during which particles are lost to collision with the planet or to scattering. 
Figure~\ref{fig:tevoall} also shows that scattering is the dominant clearing mechanism for $\mu\geq10^{-3}$; for lower planet masses, collision with the planet is the dominant clearing mechanism.

We can define a clearing timescale, $T_{cl}$, as the time required to reach 50\% of the final survivor fraction within the cleared zone. These timescales are shown as a function of $\mu$ in Figure~\ref{fig:Tcl} 
for each of the three different cases of $R_p/R_H$.  
 Clearing timescales become longer across the entire planetary mass range as the planet fills less of its Hill radius (lower $R_p/R_H$).  We observe that, $T_{cl}(\mu)$ appears to have a broken power law shape, with a shallower slope at low masses compared to a steeper slope at high masses.  The asymptotic slopes at high and low masses are similar for the three cases of $R_p/R_H$.  The transition mass, $\mu_b$, between the shallower and steeper slopes increases roughly linearly with $R_p/R_H$; this transition mass approximately coincides with the transition between impact-dominated and scattering-dominated clearing of test particles.

\begin{table}[t]
\caption{
Best-fit double power law models for the clearing timescale.}
\label{tbl:Tcl}
\begin{center}\begin{tabular}{l c c c c}
\hline
$R_p/R_H$ & $\alpha_1$     & $\alpha_2$      & $\log_{10}T_b$ & $\log_{10}\mu_b$ \\ \hline
0.001     & $-0.35\pm$0.07 & $-1.59\pm$0.18 & 4.06$\pm$0.36 & $-3.95\pm$0.38  \\
0.01      & $-0.34\pm$0.05 & $-1.48\pm$0.19 & 3.10$\pm$0.28 & $-3.43\pm$0.35  \\ 
0.1       & $-0.37\pm$0.02 & $-1.48\pm$0.34 & 1.31$\pm$0.19 & $-2.28\pm$0.27  \\ \hline
\end{tabular}
\tablecomments{$T_b$ is given in units of the planet's orbital period.}
\end{center}
\end{table}

To describe the behavior of $T_{cl}(\mu)$, we computed a nonlinear least-squares fitting to a double power law function,
\begin{equation}\label{eq:Tcl}
T_{cl}(\mu) = {2T_b\over (\mu/\mu_b)^{-\alpha_1} + (\mu/\mu_b)^{-\alpha_2} },
\end{equation}
where $\alpha_1$ and $\alpha_2$ are the asymptotic power law slopes at $\mu\ll\mu_b$ and $\mu\gg\mu_b$, respectively, and $T_b$ is the clearing timescale at the transition mass, $\mu_b$.
The best-fit values of the parameters $T_{b}, \mu_b, \alpha_1$ and $\alpha_2$, together with their $95\%$ confidence limits, are listed in Table~\ref{tbl:Tcl}, for each of the three cases of $R_p/R_H=0.001,0.01,0.1$.
The best-fit double power law models are plotted as dashed lines in Figure~\ref{fig:Tcl}.

\section{Summary and Discussion}

Using numerical integrations, we examined the orbital evolution of particles in the chaotic zone of a planet, within the framework of the circular, planar restricted three body model, in order to determine the final destinations of chaotic particles and their clearing timescales.  Our results are summarized as follows.

\begin{enumerate}

\item For the planetary masses $\mu\lesssim10^{-5}$ and planet sizes $R_p\ge0.001R_H$, particles in initially circular orbits within the chaotic zone are lost via impact with the planet. For more massive planets, chaotic particles may either collide with the planet or scatter to large distances; scattering is the dominant fate of chaotic particles for planet-star mass ratios $\mu\gtrsim10^{-3.5}$, $10^{-3}$, and $10^{-2}$, for $R_p/R_H=0.001,0.01,0.1$, respectively.  For the entire range of parameters that we investigated, we found no particles colliding with the star.

\item  For planetary masses $10^{-9}\le\mu\le10^{-4}$, most particles with initially circular orbits in the annulus of half-width approximately one Hill radius, $R_H=(\mu/3)^{1\over3}a_p$, surrounding the planet's orbit remain stable, in low eccentricity orbits in libration about the triangular Lagrange points, L$_4$ and L$_5$.   This stable ``co-orbital'' region is approximately $20-30$\% of the radial range defined by the classical chaotic zone annulus, Eq.~\ref{eq:rowidth}. The stable co-orbital region decreases for $\mu>10^{-4}$ and vanishes for $\mu\gtrsim10^{-2}$.

\item Particles initially near the inner and outer boundaries of the chaotic zone can be long lived, surviving more than $\sim10^6$ revolutions of the planet.  A minor fraction survive on low eccentricity orbits in small pockets of stability associated with the libration centers of first order mean motion resonances.  However, most of these particles survive in eccentric orbits and spend little time in the vicinity of the planet. This long-lived chaotic population is larger for lower mass planets;  for $\mu=10^{-9}$, nearly 45\% of the outer radial extent of the chaotic zone survives in eccentric orbits for more than $10^6$ revolutions of the planet.

\item  The interior and exterior cleared zone boundaries, defined as the radius where the particle survivor fraction is reduced by more than 95\%, are not symmetric about the planet's orbit: the outer cleared zone boundary is farther from the planet than the inner cleared zone boundary.  The interior boundary  mimics the classical first-order resonance overlap zone (Eq.~\ref{eq:rowidth}), with a slightly smaller numerical coefficient, $c\simeq1.2$. The exterior boundary of the cleared zone has a slightly steeper dependence on $\mu$, and is better described by $\Delta a_{cl,ext}=1.7\mu^{0.31}a_p$.  

\item
For fixed ratio of the planet radius to its Hill radius, the timescale, $T_{cl}(\mu)$, to clear the chaotic zone has double power law behavior (Fig.~\ref{fig:Tcl}, Eq.~\ref{eq:Tcl}, Table~\ref{tbl:Tcl}).  A shallow slope prevails at small $\mu$, a steeper slope at large $\mu$.   The asymptotic behavior is approximately $T_{cl}\propto\mu^{-{1\over3}}$ for small $\mu$ and $T_{cl}\propto\mu^{-{3\over2}}$ for large $\mu$.
The shallow slope behavior correlates with particle clearing dominated by collisions with the planet; the steeper slope behavior correlates with scattering being the dominant particle loss mechanism.  For $R_p/R_H=0.001$, the transition between the two power laws occurs at $\mu_b\approx10^{-4}$; the corresponding clearing timescale is approximately $10^4$ planet revolutions.  The transition mass, $\mu_b$, increases roughly linearly with $R_p/R_H$. 

\end{enumerate}

The lack of collisions with the star is rather surprising, at first sight.  It may be due to a hidden dynamical constraint or may be owed to a computational limitation in our work.  This is an interesting topic for a future investigation.

Our results show that collisions with the planet become less frequent while scattering becomes more frequent when the planet's mass exceeds a critical mass, $\mu_b$.   The relative probabilities of particle loss by collision with the planet or by gravitational scattering are not scale-free, and $\mu_b$ depends upon the planet size.  In our work, we scaled the planet size to its Hill radius, $R_p/R_H$, and computed the relative probabilities for values of this parameter spanning a fair  range of physically relevant values.  Smaller values of this parameter greatly challenge our computational resources.  However, our results indicate that, for $R_p/R_H\longrightarrow0$, scattering is the dominant outcome.  In this limit, the clearing timescale is nearly independent of $R_p/R_H$ (as witnessed by the convergence at $\mu\gg\mu_b$ in ~Fig.~\ref{fig:Tcl}).  Using the results for $R_p=0.001R_H$ as a proxy, the clearing timescale in the limit of vanishing $R_p$ can be approximated as $T_{cl}(\mu) \approx 0.024_{-0.014}^{+0.032}\mu^{-{3\over2}}$ (in units of the planet's orbit period). 
Furthermore, in the parameter regime where clearing is dominated by collisions with the planet, the clearing timescale can be approximated as $T_{cl}\approx1.33_{-0.20}^{+0.24} \mu^{-{1\over3}}(R_p/R_H)^{-1}\approx a_p/R_p$.  Thus, collisions dominate when $R_p\gg (11\mu)^{3\over2}a_p$ for $\mu\ll1$.  We can summarize these estimates for the clearing timescale (in units of the planet's orbital period) as follows,
\begin{equation}
T_{cl} \approx 
\begin{cases} 0.024\mu^{-{3\over2}} & \text{if $R_p\longrightarrow0$ (scattering dominated),}\\
{a_p\over R_p} & \text{if $R_p\gg(11\mu)^{3\over2}a_p$, $\mu\ll1$ (collision dominated).}
\end{cases}
\end{equation}

{We remind the reader that these results are obtained for the planar case.   The three dimensional case has two additional free parameters (inclination and nodal angles) and is computationally much more demanding: particle clearing timescales will, in general, be longer, and we can expect the collision dominated clearing time to scale as $\sim R_p^{-2}$.   We hope to use the present results to develop intuition and to stimulate further numerical and theoretical analysis of this problem, so as to better understand the aggregate chaotic evolution of particles in the vicinity of a planet.}

Despite their limitations, our results indicate that particle clearing timescales can be quite long in a planet's chaotic zone,  for all but the very massive close-in planets. {Distant, high mass planets clear material predominantly by scattering rather than by collisions with the planet. }  To illustrate, Neptune would require more than $\sim20$~Myr to clear its chaotic zone at its current semi-major axis (30 AU). 
Such long clearing timescales are pertinent for the interpretation of recent observations of debris disks of young exo-planetary systems. The clearing timescale for putative long period planets sculpting these disks can be comparable to the age of the system. Therefore, given the age of a system with a detectable gap in a circumstellar disk, the chaotic zone clearing timescale can be used to derive a lower mass limit of a (usually undetected) planet required to clear the gap. 

HR 8799 and HD 95086 are two recently detected exo-planetary systems, each hosting a warm inner debris disk and a cold outer disk with inner edges beyond $\sim95$ AU and $\sim63$ AU, respectively \citep{Su:2009, Moor:2013}. {A putative planet shepherding the inner edge of the cold disk would have orbital period $\sim 750$~yr and $\sim400$~yr, respectively, and $R_p/R_H\lesssim10^{-4}$.}  For the system ages of $\sim$30~Myr and 17 Myr \citep{Zuckerman:2011, Pecaut:2012}, our results imply that a shepherding planet that could clear dust and planetesimals near the inner boundary of the cold disk must have a mass $\gtrsim10M_\oplus$.  {(Our model uses massless disk particles and a fixed orbit of the planet; if the debris disk is massive, it could significantly affect the planet's orbit, which may affect the robustness of this planet mass limit.)}  Indeed, HR 8799 is known to host four planets (directly imaged), each of  5--7~$M_{Jup}$ (based on planetary thermal models);   HD 95086 is known to host a single planet (directly imaged) interior to its cold debris disk, with an estimated mass of $5M_{Jup}$ also based on planetary thermal models~\citep{Marois:2008, Marois:2010, Rameau:2013}. The lower mass limit imposed by dynamical clearing timescales offers an independent constraint on the mass of directly imaged planets in debris disk systems; and it may be used to estimate the masses of undetected planets. This constraint may become of greater value in the future, when higher sensitivity and higher spatial resolution observations of exo-planetary systems hosting {fainter debris disks sculpted by} lower mass planets become available.

\section{Acknowledgments} 
We thank Scott Tremaine and Roman Rafikov for helpful discussions. {We also thank an anonymous reviewer for comments that have improved this manuscript.}  This research was partially supported by NASA (NESSF grant NNX13AO65H) and NSF (grant AST-1312498).   This research made use of the NASA Astrophysics Data System Bibliographic Services and the NASA Exoplanet Archive.  

\bibliographystyle{apj}

\end{document}